\documentclass[onecolumn,submit]{scias} 

\usepackage{lineno}

\FrontMatterExtraVskip{0.0pt} 

\usepackage{setspace}

\title{Interplay Between Single-Photon Ionization and the Auger Process in Argon Ion Formation}

\author{LINHAO XIONG} 
\address{Department of Physics and Astronomy, University College London, Gower Street, London WC1E 6BT, UK\\}

\ead{linhao.xiong.22@ucl.ac.uk} 

\abstract{We explore the interactions between Argon and extreme ultraviolet (XUV) laser pulses across photon energies of 200 eV, 260 eV, and 315 eV, scrutinizing the influence of photon energy on Argon ion yields and unraveling the associated ionization pathways. Utilizing pulse durations of 10 fs and 30 fs, we spotlight a notable increase in Argon's ionization propensity with escalating laser intensities, especially within the \(5 \times 10^{14}\, \text{W/cm}^2\) to \(1.2 \times 10^{16}\, \text{W/cm}^2\) range. While direct ionization predominated at 200 eV, the 260 eV and 315 eV spectra revealed complex interactions, prominently featuring the Auger process. A consistent overshadowing of odd-charged ions by even-charged Argon ions, particularly at 315 eV, paves the way for future research. The findings provide crucial insights into atomic interactions with intense light, establishing a groundwork for further exploration into similar atomic systems.}

\keywords{Argon Ionization, Pathways}

\begin{document}
\frontmatter

\section{Introduction}
The exploration of atomic interactions, particularly concerning ionization processes under the influence of laser pulses \cite{freeman1991investigations,jurek2004dynamics}, has been a focal point in recent scientific inquiries. A critical aspect of this exploration involves unraveling the complexities of direct multiphoton multi-ionization and sequential ionization, which encompasses a detailed understanding of the intricate dynamics between photoionization and Auger transitions \cite{ho2014theoretical,liu2018practical}. Prior research has delved deeply into these phenomena, seeking to elucidate the underlying mechanisms and their implications in various atomic contexts \cite{lambropoulos2011route,luk1983anomalous,doumy2011nonlinear,young2010femtosecond,sorokin2007photoelectric}.

Moreover, the application of Free Electron Lasers (FELs) has emerged as a pivotal tool in this domain, enabling researchers to capture Auger spectra \cite{huttula2013decay,lablanquie2007multielectron,pulkkinen1996correlation} and thereby, providing a window into the subtle and complex processes involved in atomic ionization \cite{wallis2014auger,lagutin2019auger,della2015one, gryzlova2019two}. While these studies have significantly advanced our understanding, a comprehensive exploration of the relationship between photon energy and ion yields\cite{li2022effects}, especially in the context of Argon interacting with extreme ultraviolet (XUV) laser pulses \cite{moshammer2007few,li2022effects}, warrants further investigation .

We seek to hone in on the interactions between Argon and XUV laser pulses, with a primary aim to discern the influence of photon energy on ion yields and to elucidate the distinct pathways contributing to each ion's formation \cite{wallis2015traces,liu2014auger}. This entails a comprehensive analysis of the sequential processes of single photon absorption and Auger transitions in ion genesis, navigating through the nuanced dynamics and providing a detailed account of the ionization processes across different photon energy levels.

The insights derived from this study underscore the profound influence of laser parameters on ionization results, suggesting that meticulous calibration of these elements is paramount in experimental design and execution. By delving into the specific interactions and ionization pathways of Argon under varied photon energies and laser intensities, this research not only contributes to the existing body of knowledge but also paves the way for further investigations into analogous atomic systems and interactions, thereby enhancing our understanding and control of atomic ionization processes in experimental physics.

\section{Methodology}
For a deeper insight into how atoms react to free-electron laser pulses and the intricate relationship between photo-ionization and Auger decays, rate equations \cite{wallis2014auger,makris2009theory} can be formulated and resolved. These mathematical representations facilitate the determination of the population for every electronic configuration and reachable ion state. Moreover, they depict how these populations evolve with time, influenced by processes like photo-ionization and Auger decay.

In the realm of spectroscopy, the term "states" refers to the various electronic configurations that an atom or molecule can adopt during a given process \cite{mcnaught1997compendium}. These configurations can be represented by a sequence of numbers denoting the electrons occupying each energy level. 

In order to understand the dynamics of these states during a laser pulse, rate equations can be used to calculate the population of each configuration over time. By taking into account these processes after the laser pulse ends, we can accurately determine the decay of electronic configurations to their stable states.
The rate equations are given below \cite{wallis2015traces,banks2019exotic}:
\begin{equation}
\begin{aligned}
    \frac{d}{dt}\mathcal{I}_{j}^{(q)}(t) &= \sum_i{\left[ \sigma _{i\rightarrow j}J(t) +\Gamma _{i\rightarrow j} \right]}\mathcal{I}_{i}^{(q-1)}(t) \\
    &\phantom{=} - \sum_k{\left[ \sigma _{j\rightarrow k}J(t) +\Gamma _{j\rightarrow k} \right]}\mathcal{I}_{j}^{(q)}(t), \\
    \frac{d}{dt}\mathcal{A}_{i\rightarrow j}^{(q)} &= \Gamma _{i\rightarrow j}\mathcal{I}_{i}^{(q-1)}(t), \\
    \frac{d}{dt}\mathcal{P}_{i\rightarrow j}^{(q)} &= \sigma _{i\rightarrow j}J(t)\mathcal{I}_{i}^{(q-1)}(t).
\end{aligned}
\end{equation}

Here \(\mathcal{I}_{j}^{(q)}(t)\) is the population of an ion state \(j\) with charge \(q\), \(J(t)\) is the photon flux \cite{wallis2014auger,rohringer2007x}. \(\mathcal{A}_{i\rightarrow j}^{(q)}\) is the Auger electron yield and \(\mathcal{P}_{i\rightarrow j}^{(q)}\) is the photo-electron yield. These quantities indicate the probability of detecting a photo-electron or an Auger electron, respectively \cite{wallis2014auger,banks2019exotic}.

The parameter \(\sigma_{i\rightarrow j}\) quantifies the single-photon absorption cross-section for the electron's transition from the bound orbital \(i\) to the continuum orbital \(j\). This transition probability is integral to understanding the electron dynamics. The cross-section is calculated using the following equation \cite{sakurai1995modern,edmonds1996angular}:
\begin{equation}
\sigma_{i\rightarrow j}=\frac{4}{3}\alpha \pi ^2\omega N_i\sum_{lm}\sum_{M=-1,0,1}\left| D_{i\rightarrow j}^{M} \right|^2
\end{equation}
In this equation, \(\alpha\) represents the fine structure constant, illustrating the strength of the electromagnetic interaction. \(N_i\) is the number of electrons in the initial state orbital \(i\), \(\omega\) denotes the photon energy, \(D_{i\rightarrow j}^{M}\) is the couple part which can be expressed in matrix element and \(M\) specifies the photon's polarization \cite{banks2019exotic, sakurai1995modern,edmonds1996angular}.

The decay rate \(\Gamma_{i\rightarrow j}\) of the Auger process is derived using Fermi’s Golden Rule, which provides a framework for calculating transition probabilities in quantum systems \cite{banks2019exotic,pauli2000wave}. The decay rate for the transition involving two initial electron states \(n_1l_1\) and \(n_2l_2\), and a final state \(n_3l_3\), is given by \cite{shevelko2012atomic,pauli2000wave}:
\begin{equation}
\Gamma _{i\rightarrow j}\left( n_1l_1,n_2l_2\rightarrow n_3l_3 \right) = 2\pi N_{12}\sum{|<j|H'|i>|^2} = \frac{\pi N_{12}}{2l_3+1}\sum_{LSJM}\sum{|<j|H'|i>|}^2
\end{equation}
In this equation, \(H'\) represents the interaction Hamiltonian, and \(N_{12}\) is a weighting factor accounting for the initial electron states. This formulation assumes the two electrons are initially in states \(n_1 l_1\) and \(n_2 l_2\), with one electron transitioning to fill in state \(n_3 l_3\).

To ascertain the available pathways, we initiate by scrutinizing the initial state of the neutral atom and identifying all conceivable allowed transitions emanating from this state. Subsequently, we generate a comprehensive set of pathways by appending a new available pathway corresponding to each of these transitions. In executing this, we adopt an iterative approach, wherein a new pathway is created for every permissible transition from the final state of each pre-existing pathway. This process perpetuates until no additional transitions are plausible for any pathway \cite{banks2019exotic}.

In light of this, the rate equation necessitates a corresponding rewrite \cite{banks2019exotic}:
\begin{equation}
\begin{aligned}
    \frac{d}{dt}\mathcal{I}_{i\rightarrow j\rightarrow k}^{(q)}(t) 
    &= \left[ \sigma_{j\rightarrow k}J(t) + \Gamma_{j\rightarrow k} \right] \mathcal{I}_{i\rightarrow j}^{(q-1)}(t) \\
    &\phantom{=} - \sum_l \left[ \sigma_{k\rightarrow l}J(t) + \Gamma_{k\rightarrow l} \right]\mathcal{I}_{i\rightarrow j\rightarrow k}^{(q)}(t),
\end{aligned}
\end{equation}

Here \(\mathcal{I}_{i\rightarrow j\rightarrow k}^{\left( q \right)}\left( t \right)\) is the population of an ion state \(k\) with charge \(q\) in the pathway \(i\rightarrow j\rightarrow k\) and \(\mathcal{I}_{i\rightarrow j}^{\left( q-1 \right)}\left( t \right)\) is the population of an ion state \(j\) with charge \(q-1\) in the pathway \(i\rightarrow j\).  These cross-sections are calculated by atomic physics codes\cite{lanl_atomic_codes} while the Auger decay rates are performed with quantum chemistry package\cite{werner2009molpro}.

\section{Results and Discussion}
\subsection{Laser Pulse of 200 eV}
A meticulous examination of the correlation between state yields and the number of charges is pivotal, providing crucial insights into the distribution of Argon's various states. Notably, Argon ions, even with identical charges, can manifest different electronic configurations. For instance, an \( \text{Ar}^{+} \) ion may lose an electron from the \( 2s \), \( 2p \), \( 3s \), or \( 3p \) energy levels, contingent upon specific conditions. Acknowledging these variations is imperative for identifying the primary electron loss pathways, which will be further explored in subsequent sections.

\begin{figure}[htbp]
    \centering
    \includegraphics[width=13cm]{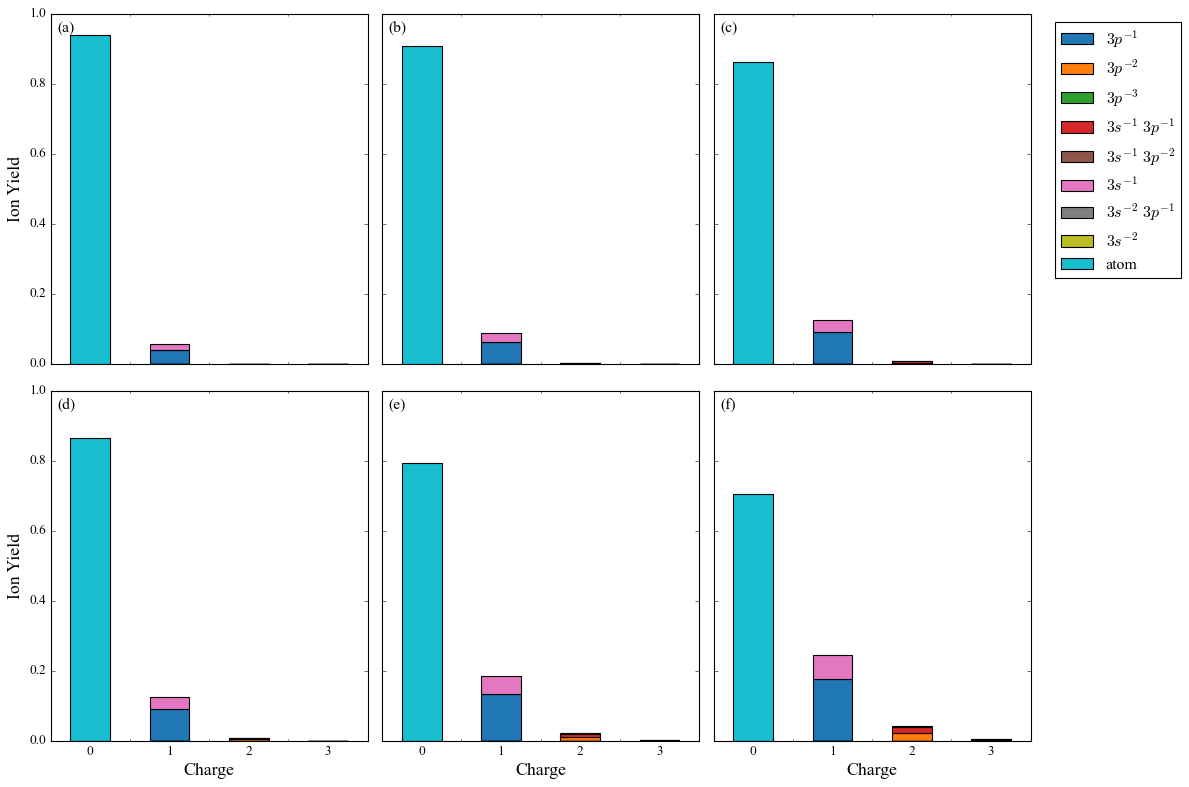}
    \caption{Distribution of Argon State Yields at 200 eV Photon Energy: A comparison between pulse durations of 10 fs (a-c) and 30 fs (d-f) at intensities of \(5 \times 10^{14}\) W/cm\(^2\) (a,d), \(8 \times 10^{14}\) W/cm\(^2\) (b,e) and \(1.2 \times 10^{15}\) W/cm\(^2\) (c,f). For clarity, labels such as \(3s^{-a} 3p^{-b}\) are used to represent the electronic configuration \(3s^{2-a} 3p^{6-b}\).}
    \label{fig:state 200 eV}
\end{figure}

From the three sets of plots presented in Fig.\ref{fig:state 200 eV}, a direct relationship between the laser's intensity, pulse duration, and the state yields of Argon ions is palpably evident. An escalation in laser intensity amplifies the energy imparted to the Argon atoms, enhancing ionization likelihood. Simultaneously, a prolonged pulse duration subjects the atoms to the laser for an extended period, further facilitating ionization. Consequently, under these conditions, the state yields for higher Argon ions also experience a notable elevation.

For both \(10 \, \text{fs}\) and \(30 \, \text{fs}\) pulse durations, within an intensity range of \(5 \times 10^{14} \, \text{W/cm}^2\) to \(1.2 \times 10^{15} \, \text{W/cm}^2\), the final ion state can only lose electrons where \(n=3\), and it is primarily characterized by the \(3p^{-b}\) electronic configuration.

\begin{figure}[htbp]
    \centering
    \includegraphics[width=15cm]{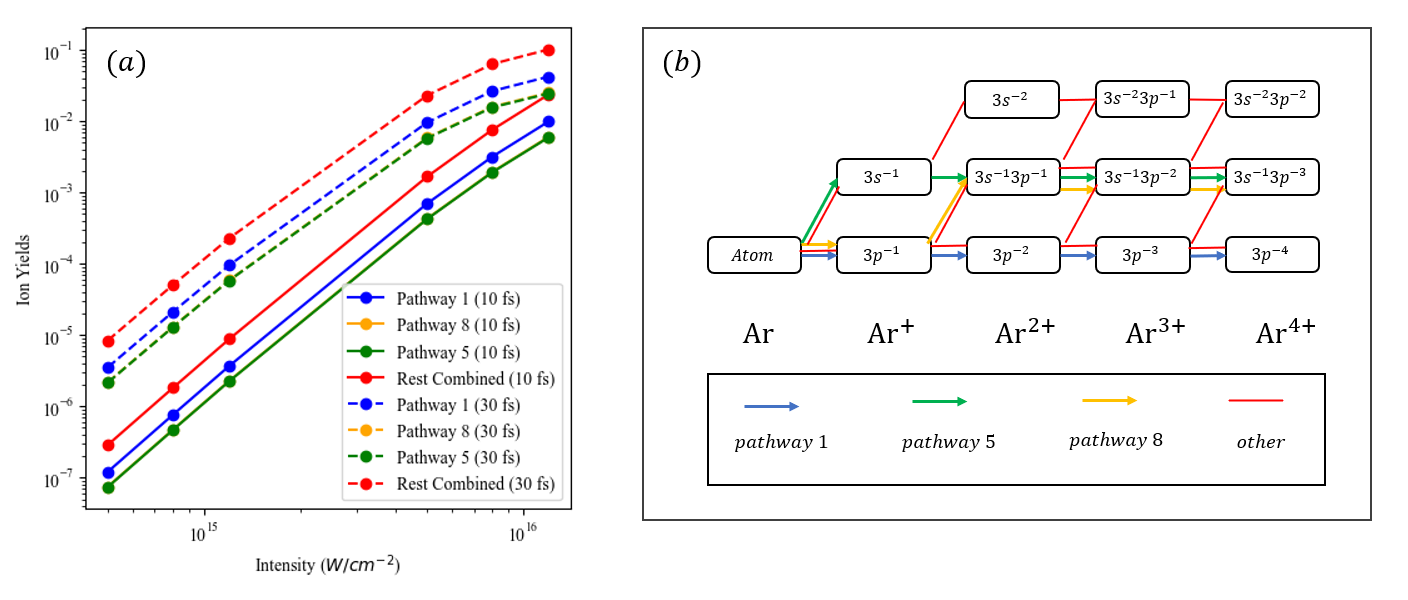}
    \caption{(a) Ion Yields for \( \text{Ar}^{4+} \) via different pathways at 200 eV: pulse duration 10 fs (solid lines) and 30 fs (dashed lines). Note: pathway 5 and pathway 8 overlapped. (b) In Pathway 1, Argon sheds electrons from its outer \(3p\) shell, transitioning from the \(3s^2 3p^6\) configuration to \(3s^2 3p^2\). Distinctly, Pathway 8 follows a two-step procedure: it begins with ionization from the deeper \(3s\) orbital and then undergoes successive ionization from the \(3p\) orbitals, culminating in a \(3s^1 3p^3\) configuration. Meanwhile, Pathway 5 adopts a mixed strategy. It commences with ionization from the \(3p\) shell, then the \(3s\), and finally shifts its primary focus back to the \(3p\) shell, settling into a \(3s^1 3p^3\) configuration.}
    \label{fig:pathway200ev}
\end{figure}

An exploration into the ionization pathways of \( \text{Ar}^{4+} \) reveals intriguing consistencies and robustness in the foundational ionization dynamics of Argon, particularly evident across pulse durations of 10 fs and 30 fs at an energy level of 200 eV. The primary ionization routes, despite the variation in pulse duration, remain notably unaffected, underscoring a stability in the ionization processes under investigation. Three primary pathways, designated as Pathways 1, 8, and 5 in Fig.\ref{fig:pathway200ev} (a), emerge as particularly significant in the ionization of \( \text{Ar}^{4+} \), collectively accounting for nearly half of all observed ionization processes. 

These pathways, detailed in Fig.\ref{fig:pathway200ev} (b), provide a comprehensive view of the predominant ionization strategies adopted by Argon under the specified conditions.

For ion states \( \text{Ar}^{4+} \), even similarly for \( \text{Ar}^{5+} \), and \( \text{Ar}^{6+} \), ionization at this photon energy is dictated solely by a single-photon mechanism, bypassing the Auger process. This emphasizes the primacy of direct ionization at 200 eV, eschewing the intricate multi-electron dynamics that the Auger process might introduce.

\subsection{Laser Pulse of 260 eV}
Adopting a methodology analogous to that utilized at a photon energy of 200 eV, we graphically delineated the ion yield distributions for various Argon ion states, categorizing them based on their respective charges.

\begin{figure}[htbp]
    \centering
    \includegraphics[width=13cm]{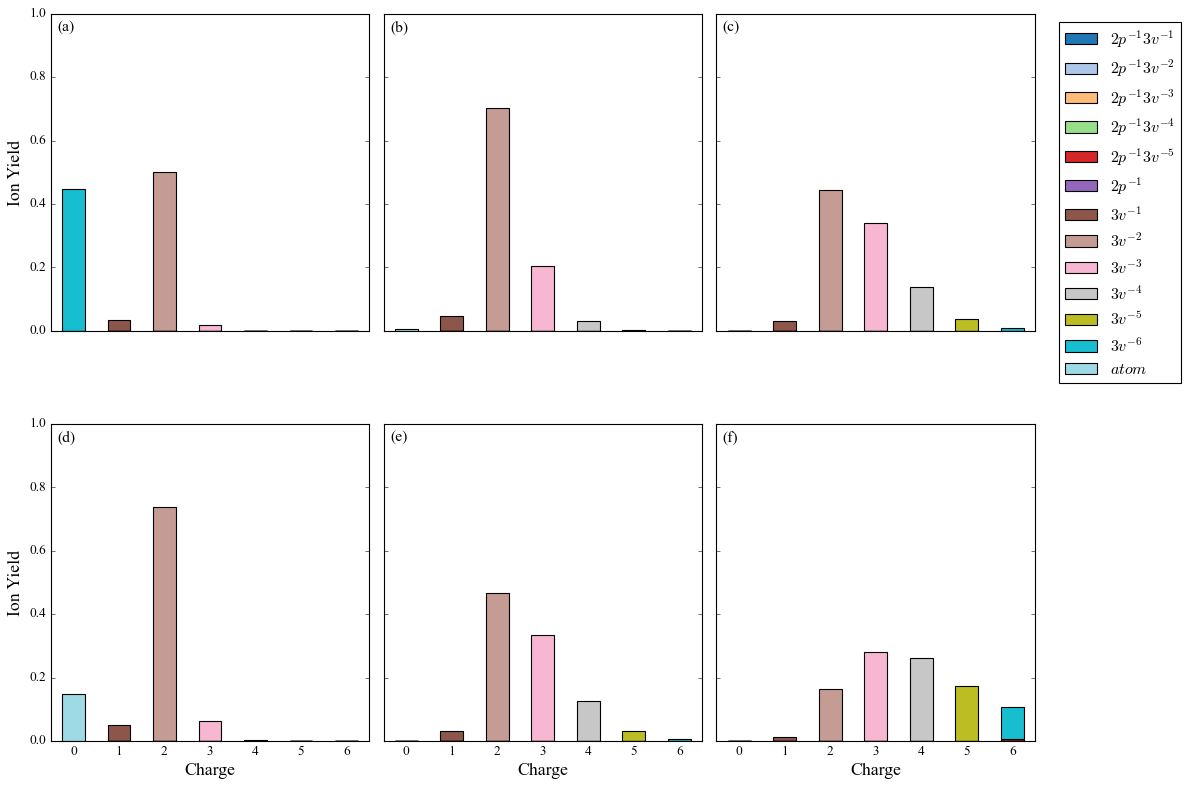}
    \caption{Distribution of Argon State Yields at 260 eV Photon Energy: A comparison between pulse durations of 10 fs (a-c) and 30 fs (d-f) at intensities of \(8 \times 10^{14}\) W/cm\(^2\) (a,d), \(5 \times 10^{15}\) W/cm\(^2\) (b,e) and \(1.2 \times 10^{16}\) W/cm\(^2\) (c,f). A new notation has been introduced for this energy level: \(2p^{-a} 3v^{-b}\), which corresponds to the electronic configuration \(1s^{2} 2s^{2} 2p^{6-a} 3s^{2-c} 3p^{6-d}\), where \(b=c+d\).}
    \label{fig:state 260 eV}
\end{figure}

At 260 eV, ion yields are notably higher across most ion states compared to the 200 eV data, given the same pulse duration and intensity (see in Fig.\ref{fig:state 260 eV}). A predominant trend emerges, showcasing an initial spike in yields with increasing intensity, which eventually plateaus or slightly declines upon reaching a peak. Enhanced laser intensity and prolonged pulse durations consistently lead to augmented yields in more highly ionized Argon species, underlining their increased sensitivity to these factors. Whether the pulse is 10 fs or 30 fs, primary ionization largely stems from electron losses at the \(n=3\) energy level. Additionally, a marked transition is observed in the ion states that release electrons from the inner \(2p\) orbital.

At a photon energy of 260 eV, the formation of Argon ions is shaped by a confluence of mechanisms, not only encompassing single photoionization but also incorporating the Auger process. This melding introduces a layer of complexity and variability into the ionization pathways, as the Auger process brings additional multi-electron dynamics into the ionization mechanisms. To streamline representation and analysis, we have categorized the ionization pathways, indicating whether electron loss occurs from the inner or outer cell.

\begin{figure}[htbp]
    \centering
    \includegraphics[width=15cm]{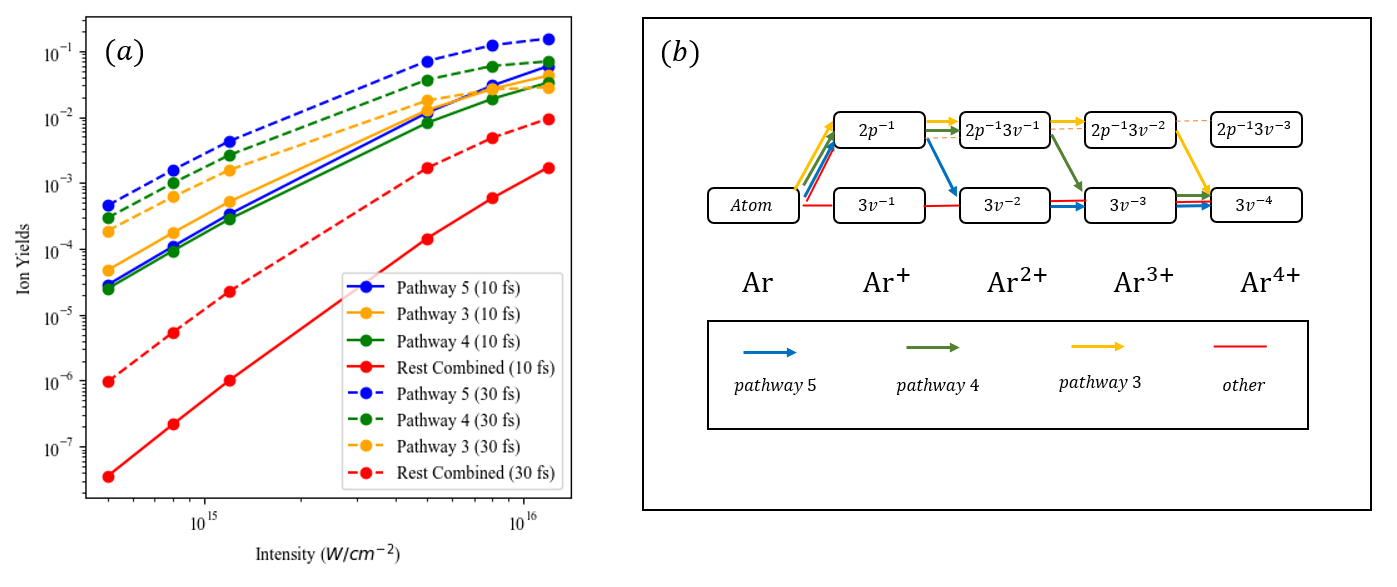}
    \caption{(a) Ion Yields for \( \text{Ar}^{4+} \) via different pathways at 260 eV: pulse duration 10 fs (solid lines) and 30 fs (dashed lines). (b) For \( \text{Ar}^{4+} \), three primary pathways, designated as 5, 3, and 4, emerge as the most significant. Together, they account for nearly all the observed ionization processes, with each involving the Auger process. Pathway 5 initiates with a \( 2p \) electron loss, followed by the Auger process, culminating in a \( 3v^{-4} \) configuration after two additional electron losses. Pathway 3 starts by shedding a \( 2p \) electron, then loses two from the \( 3v \) orbital, with a subsequent Auger process finalizing it in a \( 3v^{-4} \) state. Meanwhile, Pathway 4 begins with a \( 2p \) electron loss and, after an additional \( 3v \) electron loss and the Auger process, settles in the same \( 3v^{-4} \) configuration.}
    \label{fig:pathways260ev}
\end{figure}
Pulse duration becomes another determinant for the primary pathways (see in Fig.\ref{fig:pathways260ev} (a)), which isn't evident at 200 eV. When the pulse duration is extended from 10 fs to 30 fs in the case of Argon ions, the first three pathways remain unchanged, but there can be a rearrangement in their order of prominence.

Significantly, at 260 eV, all primary ionization pathways for higher-charged Argon ions incorporate the Auger process for both mentioned pulse durations (see in Fig.\ref{fig:pathways260ev} (b)). However, in contrast to 200 eV, the sequence of pathways at 260 eV is influenced by pulse duration: shorter pulses typically culminate with the Auger process, while longer ones invoke it earlier, immediately following the detection of an inner cell vacancy.

\subsection{Laser Pulse of 315 eV}
\begin{figure}[htbp]
    \centering
    \includegraphics[width=13cm]{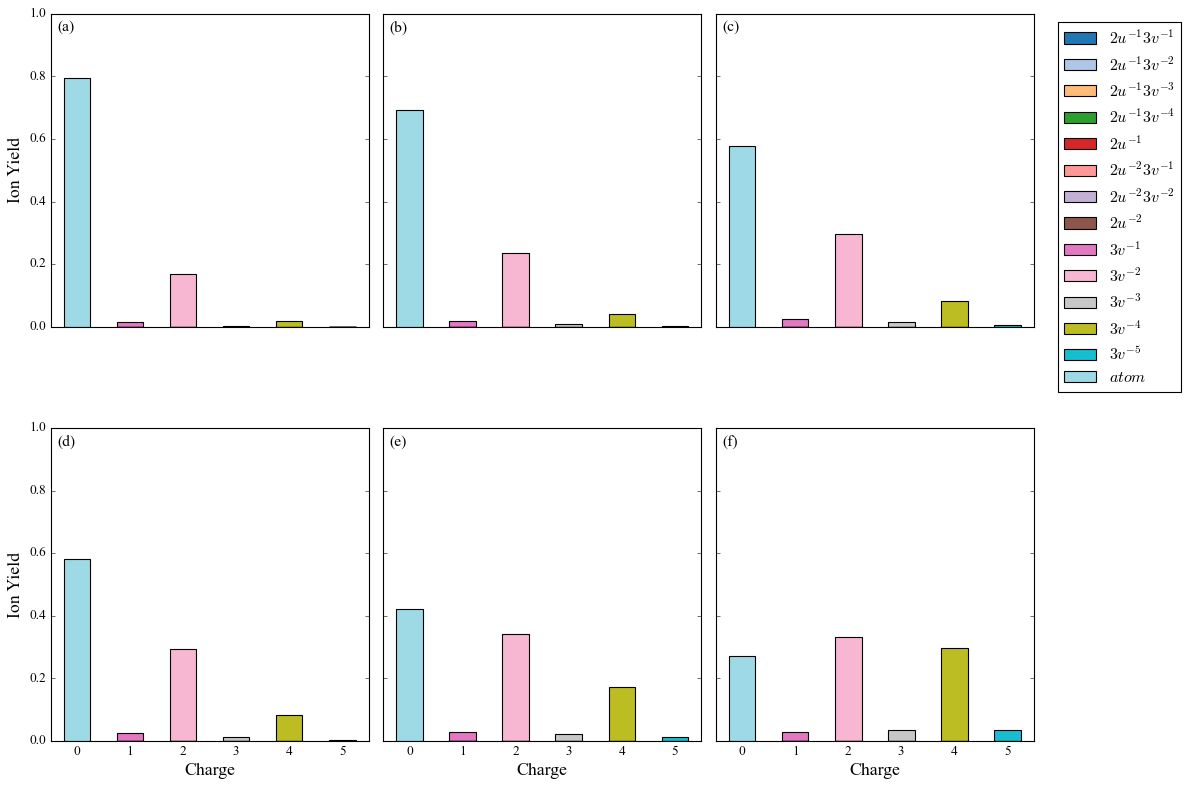}
    \caption{Distribution of Argon State Yields at 315 eV Photon Energy: A comparison between pulse durations of 10 fs (a-c) and 30 fs (d-f) at intensities of \(5 \times 10^{14} \, \text{W/cm}^2\) (a,d), \(8 \times 10^{14} \, \text{W/cm}^2\) (b,e), and \(1.2 \times 10^{15} \, \text{W/cm}^2\) (c,f). This notation, denoted as \(2u^{-a} 3v^{-b}\), streamlines the representation of the electronic configuration, which in the conventional expression is denoted as \(1s^2 2s^{2-e} 2p^{6-f} 3s^{2-c} 3p^{6-d}\). In this innovative notation, the variables 'a' and 'b' are composite values derived as \(a=e+f\) and \(b=c+d\).
}
    \label{fig:state 315 eV}
\end{figure}

At a photon energy of \(315 \, \text{eV}\), spanning laser intensities from \(5 \times 10^{14} \, \text{W/cm}^2\) to \(1.2 \times 10^{15} \, \text{W/cm}^2\), Argon ionization exhibits distinct patterns (see in Fig.\ref{fig:state 315 eV}). With escalating laser intensity and pulse duration, state yields for highly ionized Argon species consistently surge. Notably, even-charged Argon ions consistently outperform their odd-charged counterparts in yield, a phenomenon likely resulting from a complex interplay between the Auger process and sequential ionization.

\begin{figure}[htbp]
    \centering
    \includegraphics[width=15cm]{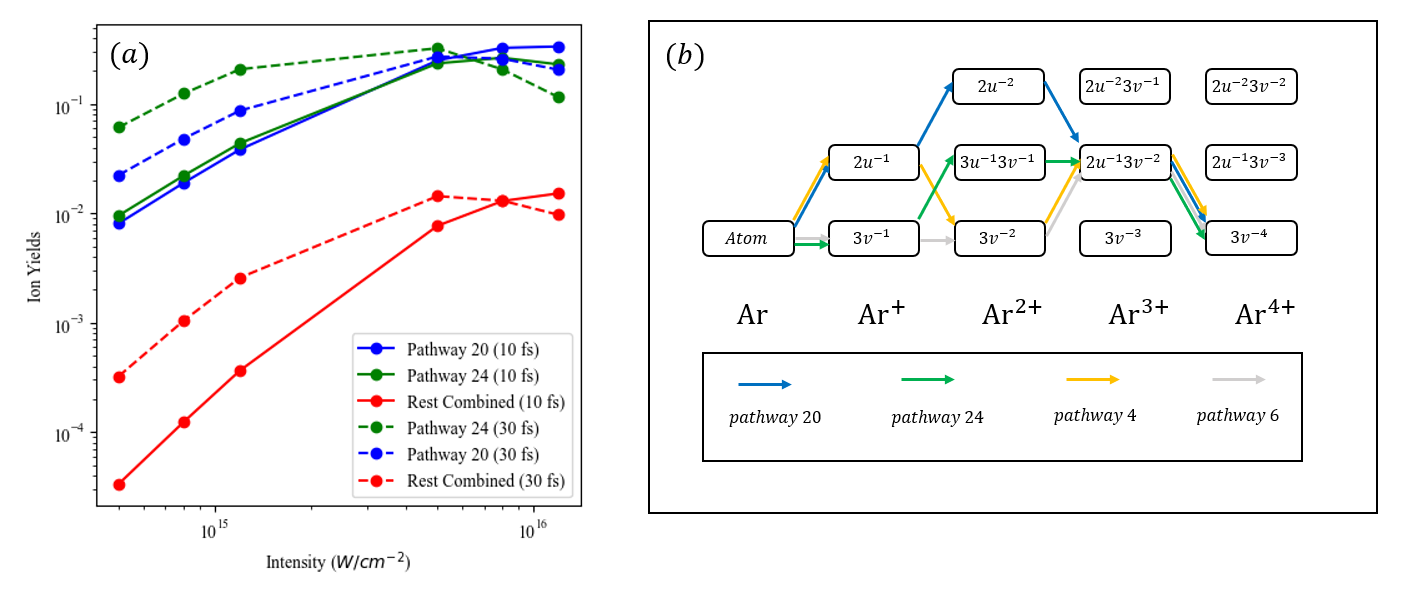}
    \caption{(a) Ion Yields for \( \text{Ar}^{4+} \) via different pathways at \( 315 \, \text{eV} \): pulse duration \( 10 \, \text{fs} \) (dashed lines) and \( 30 \, \text{fs} \) (solid lines). (b) For \( \text{Ar}^{4+} \), pathways 20 and 24 emerge as the predominant ones, collectively accounting for the majority of the ionization processes observed. Both prominently feature the Auger process. Pathway 20 begins with the sequential loss of two \(2u\) electrons, followed by two consecutive Auger processes, resulting in a \( 3v^{-4} \) configuration (\(PPAA\) sequence). In contrast, Pathway 24 initiates with the removal of a \(2u\) electron, succeeded by an Auger process that fills the \(2u\) slot and expels two \(3v\) electrons. Subsequently, another \(2u\) electron is lost, with a subsequent Auger process culminating in a \( 3v^{-4} \) state (double \(PA\) sequence). Regardless of the pulse duration, whether 10 fs or 30 fs, these pathways remain dominant.}
    \label{fig:pathways315ev1}
\end{figure}

\begin{figure}[htbp]
    \centering
    \includegraphics[width=15cm]{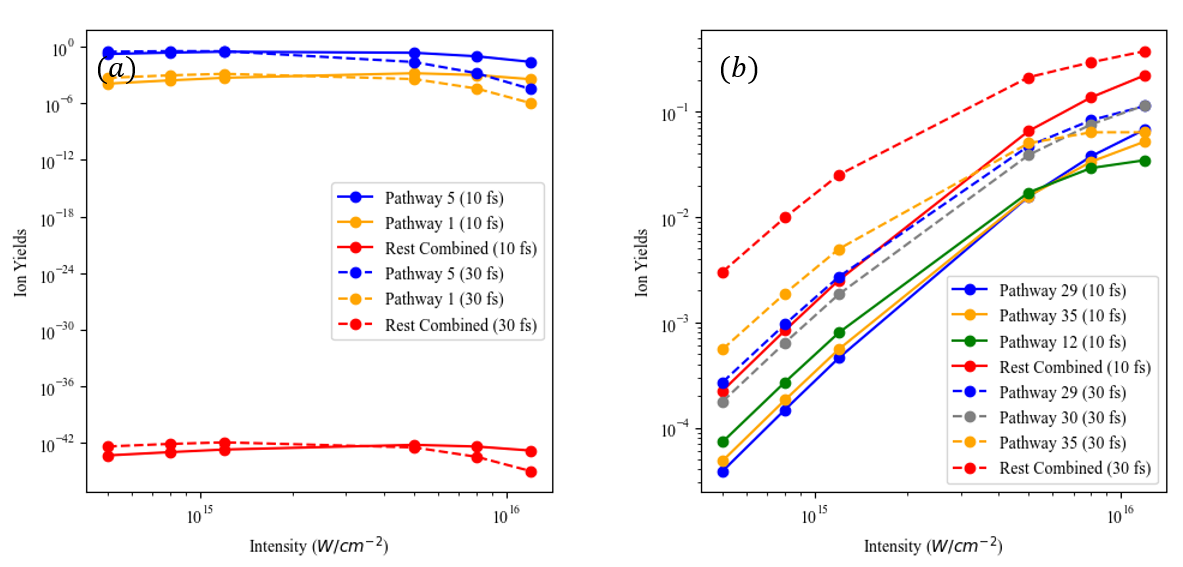}
    \caption{(a) Ion Yields for \( \text{Ar}^{2+} \) via different pathways at \( 315 \, \text{eV} \): pulse duration \( 10 \, \text{fs} \) (solid lines) and \( 30 \, \text{fs} \) (dashed lines). For the formation of \( \text{Ar}^{2+} \), the primary Pathway 5 involves single photoionization in the inner \( 2u \) cell, initially leading to an electron configuration of \( 2u^{-1} \), followed by an Auger process resulting in the final state of \( 3v^{-2} \). A secondary significant pathway, Pathway 1, involves sequential ionization from the outer cell. Together, these two pathways account for nearly all the observed ionization processes. (b)Ion Yields for \( \text{Ar}^{5+} \) via different pathways at 315 eV: pulse duration 10 fs (solid lines) and 30 fs (dashed lines)}
    \label{fig:pathways315ev2}
\end{figure}

Delving into the primary pathways under a photon energy of \(315 \, \text{eV}\). Employing the established approach, we graphically represented the ion yields corresponding to each pathway of \(\text{Ar}^{4+}\) firstly (see in Fig.\ref{fig:pathways315ev1} (a)). The dominance observed in the photoionization-Auger sequence (\(PA\)) and photon-photon-Auger-Auger sequence (\(PA\)) are indicative of a prevailing trend in even-charged ion yields. To examine this phenomenon more thoroughly, we continued testing specifically with \(\text{Ar}^{2+}\) and \(\text{Ar}^{5+}\)(see in Fig.\ref{fig:pathways315ev2}).

Both \( \text{Ar}^{2+} \) and \( \text{Ar}^{4+} \) demonstrate that the \(PA\) sequence and \(PPAA\) sequence are significant components in the pathway of Argon ion formation, particularly for even charges (see also in Fig.\ref{fig:pathways315ev1} (b)). Conversely, for \( \text{Ar}^{5+} \), pathways involving \(PA\) or \(PPAA\) sequences do not dominate; in fact, no dominating pathway can be seen in Fig.\ref{fig:pathways315ev2} (b). These facts collectively support the conclusion that these two sequences account for the observed dominance of even-charged ion yields at 315 eV photon energy.

The reason that the \(PA\) and \(PPAA\) sequences can easily occur is linked to the photon energy of 315 eV, where electrons in both the \(2p\) and \(2s\) orbitals are more readily ionized. This ionization creates holes in the inner cell, forming an ideal precondition for the subsequent Auger process. the photon ionization cross-section plays a vital role in this hole-creating process. We calculated that the cross-section of inner cell electrons is higher than that of the outer valence electrons. This discrepancy indicates that inner cell electrons have a larger possibility of ionization, as long as they satisfy the ionization's minimum energy requirement.

\section{Conclusion}
We explored the effects of photon energy on Argon ion yields, navigating through the intricate pathways behind each ion's formation with a spotlight on single-photon absorption and Auger transitions. At \(200 \, \text{eV}\), Argon’s susceptibility to ionization notably surged with increased laser intensity, particularly within the range of \(5 \times 10^{14} \, \text{W/cm}^2\)  to \(1.2 \times 10^{16} \, \text{W/cm}^2\) with ion states up to \(\text{Ar}^{6+}\) predominantly influenced by single-photon mechanisms, thereby sidelining the Auger process. Transitioning to \(260 \, \text{eV}\), a discernible increase in ion yields across most states was observed, with heightened ionization sensitivity notably tethered to the Auger process, and its sequence manifesting variations with pulse duration. At \(315 \, \text{eV}\), Argon exhibited a pronounced upsurge in ionization, surpassing the yields at \(200 \, \text{eV}\) and \(260 \, \text{eV}\), and introducing a consistent pattern of even-charged Argon ions eclipsing their odd-charged counterparts.

The insights gleaned illuminate the profound influence of laser parameters on ionization results, emphasizing the paramount importance of meticulous calibration in experimental design. Future investigations will delve deeper into this phenomenon, focusing on understanding the specific loss of electron orbitals with direction, such as the \(2p_x\) and \(2p_y\) orbitals, thereby paving the way for further explorations in the expansive domain of atomic and molecular physics.
\acknowledgements{I would like to express my deepest gratitude to Prof. Agapi Emmanouilidou for her unwavering guidance, support, and patience throughout my Master of Science program. Her insights and expertise have been invaluable to my academic growth and the completion of this project. Special thanks go to Mr. Miles Mountney, a PhD student from our research group, for his invaluable assistance in running computational codes and data collection. His contributions were essential in bringing this research to fruition.} 

\bibliography{scias_template}

\end{document}